\documentclass[aps,prl,twocolumn,superscriptaddress,showpacs,floatfix]{revtex4}
\usepackage{graphicx,epsfig,amsmath}

\newcommand{\etal}{\textit{et al.}}

\newcommand{\subfig}[2]{Fig.~\ref{fig:#1}(#2)}
\newcommand{\totfig}[1]{Fig.~\ref{fig:#1}}
\newcommand{\eqn}[1]{Eq.~\ref{#1}}
\begin{document}


\title{Energy Dependent Tunneling in a Quantum Dot}

\author{K. MacLean}
    \email{kmaclean@mit.edu}
    \affiliation{Department of Physics, Massachusetts Institute of Technology, Cambridge, Massachusetts 02139}
\author{S. Amasha}
    \affiliation{Department of Physics, Massachusetts Institute of Technology, Cambridge, Massachusetts 02139}
\author{Iuliana P. Radu}
    \affiliation{Department of Physics, Massachusetts Institute of Technology, Cambridge, Massachusetts 02139}
\author{D. M. Zumb\"{u}hl}
    \affiliation{Department of Physics, Massachusetts Institute of Technology, Cambridge, Massachusetts 02139}
    \affiliation{Department of Physics and Astronomy, University of Basel, Klingelbergstrasse 82, 4056 Basel, Switzerland}
\author{M. A. Kastner}
    \affiliation{Department of Physics, Massachusetts Institute of Technology, Cambridge, Massachusetts 02139}
\author{M. P. Hanson}
    \affiliation{Materials Department, University of California, Santa Barbara 93106-5050}
\author{A. C. Gossard}
    \affiliation{Materials Department, University of California, Santa Barbara 93106-5050}


\begin{abstract}
We present measurements of the rates for an electron to tunnel on and off a quantum dot, obtained using a quantum point
contact charge sensor. The tunnel rates show exponential dependence on drain-source bias and plunger gate voltages. The
tunneling process is shown to be elastic, and a model describing tunneling in terms of the dot energy relative to the
height of the tunnel barrier quantitatively describes the measurements.
\end{abstract}


\maketitle


Gate controlled quantum dots have been used to study a wide variety of physical phenomena, from correlated-electron physics
\cite{OrigKondo:refer}, which becomes important when the coupling of the dot to its leads is strong, to the coherence of
electron charge and spin states \cite{Coherence:references}, which can be maintained only when the coupling to the leads is
weak. Observation of these phenomena is made possible by in-situ control of the rate of tunneling $\Gamma$, coupling the
dot to its leads: $\Gamma$ can be adjusted over more than ten orders of magnitude by changing the voltages applied to the
gates that define the tunnel barriers of the quantum dot. Recently, integrated charge sensors
\cite{Field1993:NoninvasiveProbe} have made possible a variety of new investigations of $\Gamma$, probing the effect
of excited states \cite{Elzerman2004:ExcitedStates,Johnson2005:ExcitedStateQuantumDot,Gustavsson:superPoisson}, electron
number \cite{ENumber:references}, barrier symmetry \cite{Haug:CoupSymm,Ensslin:CoupSymm} and Coulomb interactions
\cite{Ensslin:CoupSymm,Fujisawa:Bidirectional}.

In this Letter, we report exponential sensitivity of the rates for electrons tunneling on ($\Gamma_{on}$) and off
($\Gamma_{off}$) a lateral GaAs quantum dot as a function of the drain-source bias $V_{ds}$ and plunger gate voltage $V_g$.
The tunnel rates are obtained by performing time resolved measurements of a quantum point contact (QPC) charge sensor
adjacent to the dot \cite{Elzerman2004:SingleShotReadOut,Lu2003:RealTimeDet,Schleser2004:TimeResolvedDet} and utilizing
pulsed-gate techniques \cite{Elzerman2004:SingleShotReadOut,FujisawaTransitions}. We show that the tunneling is dominated by elastic processes, and that the measured exponential dependence of $\Gamma$ on $V_{ds}$ and $V_g$ is in excellent quantitative agreement with a model describing tunneling in terms of the dot energy relative to the height of the tunnel barrier.

\begin{figure}
\setlength{\unitlength}{1cm}
\begin{center}
\includegraphics[width=7.0cm, keepaspectratio=true]{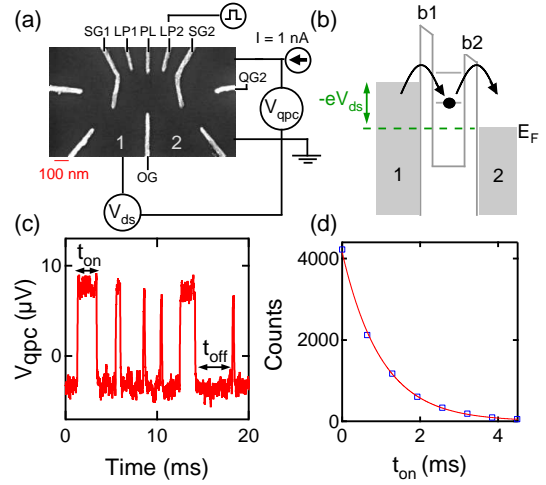}
\end{center}

\caption{(a) Electron micrograph of the gate geometry and schematic of the measurement circuit.  Unlabeled gates are
grounded. We measure the resistance of the QPC by sourcing a current through the QPC and monitoring the voltage V$_{qpc}$
across the QPC. We have verified that the finite bandwidth of our measurement \cite{FiniteBW} does not significantly affect
the results presented in this work using computer simulations. (b) When a voltage bias $V_{ds}$ is applied across the
quantum dot a small current flows and the charge on the dot fluctuates between 0 and 1 as shown in (c). We measure the time
intervals $t_{on}$ ($t_{off}$) that the electron is on (off) the dot using an automated triggering and acquisition system
described in Ref. \cite{Amasha2006:TunnelingSummary}. The offset in the trace is caused by the AC coupling of the voltage
preamplifier. (d) Histogram of $t_{on}$ times from data such as in (c).  Fitting this histogram yields $\Gamma_{off}$ as
described in the text.} \label{fig:fig1}
\end{figure}

The device used in this work is fabricated from a AlGaAs/GaAs heterostructure grown by molecular beam epitaxy. The two dimensional electron gas (2DEG) formed at the AlGaAs/GaAs interface 110 nm below the surface of the heterostructure has a mobility of 6.4 $\times$ 10$^5$
cm$^{2}$/Vs and density of 2.2 $\times$ 10$^{11}$ cm$^{-2}$ \cite{Granger2005:TwoStageKondo}.  An electron micrograph of
the gate geometry is shown in \subfig{fig1}{a} and is similar to that of Ciorga \etal \space
\cite{Ciorga2000:AdditionSpectrum}. Applying negative voltages to the labeled gates forms a single dot containing one
electron coupled to the surrounding 2DEG through two tunnel barriers: One between SG1 and OG (b1, connecting the dot to lead
1) and the other between SG2 and OG (b2, connecting the dot to lead 2). Lead 2 is grounded, and a voltage $V_{ds}$ is
applied to lead 1. Approximately the same DC voltage $V_{g}$ is applied to the three plunger gates LP1, PL, and LP2.  We
have verified that a single, not double, dot is formed.  We measure the dot in a dilution refrigerator with a lowest
electron temperature of 120 mK.

\begin{figure}

\begin{center}
\includegraphics[width=7.0cm, keepaspectratio=true]{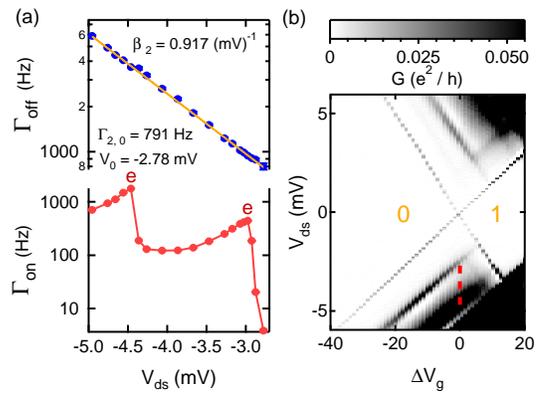}
\end{center}

    \caption{(a). $\Gamma_{on}$ and $\Gamma_{off}$ as a function of $V_{ds}$ for large negative $V_{ds}$. The solid
    line in the upper panel is based on a theoretical fit to the data discussed in the text. (b). Differential
    conductance vs. $V_{ds}$ and $V_{g}$, showing the 0 to 1 electron transition.  The tunnel rates for this case
    are made large enough so that the differential conductance can be measured using standard transport techniques.
    The data shown in (a) are taken at the position of the dashed line.}
    \label{fig:fig2}
\end{figure}

Applying a negative voltage to the gate QG2, we form a QPC between the gates QG2 and SG2.  The resistance of the QPC is
sensitive to the charge on the dot \cite{Field1993:NoninvasiveProbe}, allowing us to measure the number of electrons on the
dot in real time \cite{Elzerman2004:SingleShotReadOut,Lu2003:RealTimeDet,Schleser2004:TimeResolvedDet}.  If we apply a
drain-source bias V$_{ds}$ across the dot (\subfig{fig1}{b}), we observe the number of electrons on the dot fluctuate
between 0 and 1: A typical trace is shown in \subfig{fig1}{c}. To measure $\Gamma_{off}$, we histogram the times $t_{on}$
that the electron spends on the dot (\subfig{fig1}{d}) and fit to an exponential $Ae^{-\Gamma_{off} t_{on}}$
\cite{Schleser2004:TimeResolvedDet}. We obtain $\Gamma_{on}$ from the time intervals $t_{off}$ in the same manner. Using
this technique we measure $\Gamma_{off}$ and $\Gamma_{on}$ as a function of $V_{ds}$ (\subfig{fig2}{a}) for the case in
which there is either 0 or 1 electrons on the dot at the position shown in \subfig{fig2}{b}.  As $V_{ds}$ is made more
negative, $\Gamma_{off}$ is seen to increase exponentially.

To model the tunneling rates, we linearize the exponential in the WKB formula for tunneling through a barrier \cite{Landau} for a small
perturbation $\delta E$ to the dot energy $E$ and a small deviation $\delta U$ of the tunnel barrier potential $U(x)$:
$\Gamma \sim \Gamma_0 \exp \left[ - \left(\delta U - \delta E \right) \kappa \right]$, where $\kappa$ and $\Gamma_0$ depend
on the details of the barrier potential. In a simple capacitor model for the dot we assume a linear dependence of $\delta E$
on small perturbations $\delta V_g$ and $\delta V_{ds}$ about arbitrary but fixed $V_{ds}$ and $V_g$ values:
$\delta E = -e\alpha_{dsE} \delta V_{ds}-e\alpha_{gE} \delta V_{g}$ , where $\alpha_{dsE}$ is the ratio of the drain-source capacitance to the total dot capacitance, and $\alpha_{gE}$ is the capacitance ratio for the three plunger gates.  Similarly, $\delta U$ varies linearly as $\delta U = -e\alpha_{ds U} \delta V_{ds} - e\alpha_{g U} \delta V_{g}$, where $\alpha_{ds U}$ and $\alpha_{g U}$ give the coupling of $V_{ds}$ and $V_g$ to the barrier potential.  There will, of course, be different parameters $\alpha_{ds U}$, $\alpha_{g U}$, and $\kappa$ for the two barriers.  Note that $\Gamma$ depends exponentially on $\delta U - \delta E$, and therefore depends exponentially on $V_{ds}$ and $V_g$:  One can show that this holds independent of the particular shape $U(x)$ of the barrier potential, or the shape of the perturbation to the potential induced by the change $\delta V_g$ or $\delta V_{ds}$. Using this linearization, we can write down equations for the $V_{ds}$ dependence of $\Gamma_{off}$ and $\Gamma_{on}$, including Fermi statistics in the leads, assuming elastic tunneling, and considering only a single orbital state of the dot:
\begin{eqnarray}
&\Gamma_{off} = &\Gamma_{2,0} e^{-\beta_{2}\delta V_{ds}} (1 - f_{2}(E)) \label{eq:spinWKBoff}\\
&&+\Gamma_{1,0} e^{\beta_{1}\delta V_{ds}} (1 - f_{1}(E)) \nonumber \\
&   \Gamma_{on} = &\eta \Gamma_{2,0} e^{-\beta_{2}\delta V_{ds}}f_{2}(E) \label{eq:spinWKBon} \\
&&+ \eta \Gamma_{1,0} e^{\beta_{1}\delta V_{ds}}f_{1}(E) \nonumber
\end{eqnarray}
Here $\beta_{1,2} = \kappa_{1,2}|\alpha_{dsU_{1,2}} - \alpha_{dsE}|$, $E = -e\alpha_{dsE} V_{ds}-e\alpha_{gE} \Delta V_{g}$ is the energy of the one-electron state relative to the Fermi energy $E_{F}$, and $\Delta V_{g}$ is the shift in $V_{g}$ from the 0 to 1 electron transition. $f_{2}$ and $f_{1}$ are the Fermi functions of the two leads $f_{2}(E) = f(E)$, $f_{1}(E) = f(E + eV_{ds})$, and $\eta$ is the ratio between the tunnel rate onto and off of the dot for a given lead when the one-electron state is aligned with the Fermi level in that lead. We expect that $\eta = 2$ because of spin degeneracy \cite{Cobden:SpinSplitting,Fujisawa:Bidirectional}, and use this value in the calculations below. 

To understand \subfig{fig2}{a}, we note that whether $\Gamma$ increases or decreases with $V_{ds}$ depends on whether
lead 1 is better coupled to the barrier or the dot, that is, whether $\alpha_{dsU}$ or $\alpha_{dsE}$ is larger. Since b1
is closer to lead 1 than the dot, and b2 is farther from lead 1 than the dot, it follows that $\alpha_{dsU_{1}} >
\alpha_{dsE} > \alpha_{dsU_{2}}$ (see \subfig{fig1}{b}) \cite{Antonov:Metastable}.  Therefore, tunneling through b1 (b2)
increases (decreases) exponentially with increasing $V_{ds}$. This is reflected in the signs of the exponentials appearing in
\eqn{eq:spinWKBoff} and \eqn{eq:spinWKBon}.

The solid line through the the $\Gamma_{off}$ data in \subfig{fig2}{a} is a fit to \eqn{eq:spinWKBoff}, which for large
negative bias reduces to $\Gamma_{2,0} e^{-\beta_{2}\delta V_{ds}}$.  The rate $\Gamma_{on}$ is shown as a function of
$V_{ds}$ in the lower panel of \subfig{fig2}{a}.  At the two points marked ``e'' in the figure $\Gamma_{on}$ increases
rapidly as the Fermi energy in lead 1 is aligned with an excited orbital state of the dot \cite{Gustavsson:superPoisson},
giving the 1st and 2nd excited state energies to be 1.9 and 2.9 meV, respectively, above the ground state. The higher-energy
states are better coupled to the leads and thus $\Gamma_{on}$ rises rapidly when these states become available. These
energies can also be measured, with larger tunneling rates through b1 and b2, using standard transport techniques
\cite{Kouwenhoven1997:NatoReview} (\subfig{fig2}{b}), and the results are consistent. We note that $\Gamma_{off}$ does not
show any special features at these  points: This is because the electron decays rapidly out of the excited orbital states
and subsequently tunnels off the dot from the ground state \cite{Gustavsson:superPoisson}. We therefore continue to use
\eqn{eq:spinWKBoff} when there are multiple orbital states in the transport window.

\begin{figure}
\setlength{\unitlength}{1cm}
\begin{center}
\includegraphics[width=7.0cm, keepaspectratio=true]{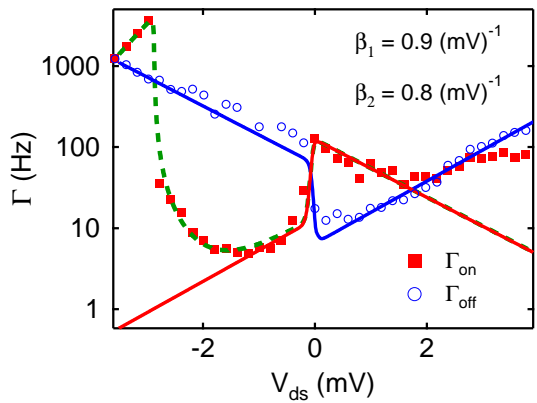}
\end{center}
\caption{$\Gamma_{on}$ (closed squares) and $\Gamma_{off}$ (open circles) as a function of $V_{ds}$.
The solid and dashed curves are calculations described in the text. The step features near $V_{ds} = 0$ result from the Fermi distribution.}
    \label{fig:fig3}
\end{figure}

In the regions between the points marked ``e'', $\Gamma_{on}$ is seen to decrease exponentially as $V_{ds}$ is made more
negative, as expected from \eqn{eq:spinWKBon}. Note that this decrease in $\Gamma_{on}$, with increasingly negative
$V_{ds}$, occurs even though the number of electrons that could tunnel onto the dot inelastically from lead 1 is growing,
presenting strong evidence that the tunneling is predominantly elastic, dominated by states very close to the dot energy
$E$.  There is, however, an apparent flattening of $\Gamma_{on}$ above the extrapolated exponential decrease near $V_{ds}$ = -4 mV, close to the second excited orbital state.  We find that this line shape is consistent with broadening of the second excited state by
a lorentzian of full-width at half-maximum $\gamma \sim 10$ $\mu$eV.  Calculated line shapes are shown for the broadening of the first excited state in \totfig{fig3} and are discussed below.

If a square barrier is assumed, one can compute an effective barrier height $U_{2}$ and width $w_{2}$ for b2 from the fit
in \subfig{fig2}{a}. For a square barrier $\Gamma = f_{0}e^{-2w_{2}\sqrt{2m^{\ast}(U_{2} - E)/\hbar^{2}}}$ \cite{Landau}.
Linearizing the square root in the exponential and estimating $\alpha_{dsE} - \alpha_{dsU_{2}}\sim \alpha_{dsE}$ and $f_{0}\sim E_{\ell s}/h \sim$ 1 THz, where $E_{\ell s}$ is the level spacing of the dot, we obtain $w_{2}\approx$\space130 nm and $U_{2} - E_{F}\approx$\space5 meV at $V_{ds} = V_{0}$. These values are only logarithmically sensitive to $f_{0}$ and thus depend very little on our estimate of this parameter.
The voltages we apply to the gates to create the tunnel barriers are the same order of magnitude as the voltages at which
the 2DEG depletes and thus it is reasonable that $U_{2} - E_{F}$ is found to be comparable to the Fermi energy ($E_{F}
\approx$ 7.7 meV). The value for $w_{2}$ is also reasonable given the dimensions of our gate pattern and heterostructure.

We next examine, in \totfig{fig3}, the dependence of $\Gamma_{on}$ and $\Gamma_{off}$ on $V_{ds}$ for both positive and
negative $V_{ds}$. The data are taken with $V_{g}$ adjusted so that $E$ is $\sim kT$ away from the 0 to 1 electron
transition at $V_{ds}$ = 0. The solid lines in \totfig{fig3} are calculated from \eqn{eq:spinWKBoff} and \eqn{eq:spinWKBon}
and are in good agreement with the data.  We note that the value of $\beta_{1}$ is very close to the value of $\beta_{2}$.
If the height and width of b1 and b2 are comparable one can show that $\beta_{1} \sim \frac{\alpha_{dsU_{1}} -
\alpha_{dsE}}{\alpha_{dsE} - \alpha_{dsU_{2}}}\beta_{2}$, and it is therefore expected that $\beta_{1} \sim \beta_{2}$.

At $V_{ds} \approx - 3$ mV in \totfig{fig3}, the first excited orbital state is aligned with the Fermi energy in lead 1. For
slightly less negative values of $V_{ds}$, where the excited state is just above the Fermi energy, $\Gamma_{on}$ appears
enhanced above the solid line calculated from \eqn{eq:spinWKBon}. The dashed line in \totfig{fig3} includes broadening 
of the first excited orbital state by a lorentzian of full-width at half-maximum $\gamma = 13$  $\mu$eV, corresponding to a lifetime broadening of $\tau_{e1} = 50 \pm 30$ ps.  Emission of acoustic phonons can lead to excited state lifetimes $\sim 100$ ps, but for our device parameters we expect much slower relaxation from this mechanism \cite{FujisawaTransitions,climentePRB,BockelmannPRB}.   $\Gamma_{on}$ also deviates from the solid curve for $V_{ds} \gtrsim 2$ mV: This deviation may be caused by broadening of the first excited state as well. These deviations could also be caused by inelastic processes, which might begin to contribute significantly to $\Gamma_{on}$ when the electron energy is sufficiently far below the Fermi energy.

\begin{figure}[!]
    \begin{center}
        \includegraphics[width=7.0cm, keepaspectratio=true]{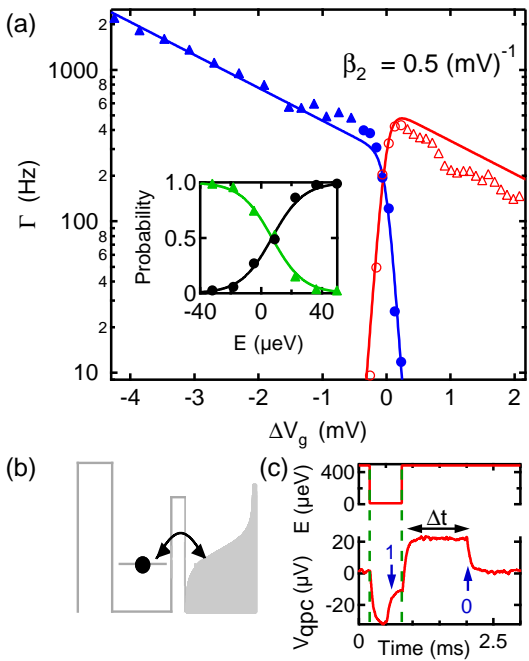}
    \end{center}
    \caption{(a) $\Gamma_{on}$ and $\Gamma_{off}$ as a function of $\Delta V_{g}$.  Closed (open) circles are
    $\Gamma_{off}$ ($\Gamma_{on}$) measured by observing spontaneous hopping caused by thermal broadening in the
    leads as depicted in (b). Closed (open) triangles are $\Gamma_{off}$ ($\Gamma_{on}$) measured using a pulsed
    gate technique. Solid lines are calculations as described in the text. (inset) $p_{on}$ (triangles) and $p_{off}$
    (circles) compared to $f(E,\eta)$ and $1 - f(E,\eta)$ (solid lines) respectively as described in the text.
(b) Sketch of spontaneous hopping caused by thermal broadening in the leads (c) Pulsed technique used to measure
$\Gamma_{off}$.  The top panel shows the pulsed modulation of the one-electron state energy $E$. The bottom panel shows a
sample time trace.  The dashed vertical lines indicate when the gate pulse begins and ends: The QPC responds to the gate
pulse because of direct capacitive coupling to LP2.  When the electron energy level is brought near $E_{F}$ an electron
tunnels on the device (indicated by a 1).  When the electron level is brought back above $E_{F}$ the electron tunnels off
the device (indicated by a 0).}
    \label{fig:fig4}
\end{figure}

We now turn to the dependence of the tunnel rates on the plunger gate voltage $V_{g}$ (see \totfig{fig4}). For these
measurements, the barriers are tuned so that the tunneling through b1 is negligible. In the region near $\Delta V_{g} = 0$
(where $E \sim$ kT), the electron hops on and off the dot spontaneously because there are both electrons and holes in the
lead at these energies (\subfig{fig4}{b}). In this region we measure the rates $\Gamma_{on}$ and $\Gamma_{off}$ in the same
way as described above. In the inset to \subfig{fig4}{a} we compare the probabilities that the electron is on and off the
dot $p_{on}=\Gamma_{on}/(\Gamma_{on} + \Gamma_{off})$ and $p_{off}=\Gamma_{off}/(\Gamma_{on} + \Gamma_{off})$ to $f(E,\eta)
= \frac{1}{1 + \frac{1}{\eta}e^{E/kT}}$ and $1 - f(E,\eta)$, respectively, with T = 120 mK \cite{Gustavsson:superPoisson}.

A pulsed technique (see \subfig{fig4}{c}) is used to measure $\Gamma_{off}$ when $\Delta V_{g}$ is made sufficiently negative
that thermally assisted electron tunneling ceases.  We begin with the electron energy, determined by $\Delta V_{g}$, well
above $E_{F}$.  We then apply a positive voltage pulse to the gate LP2 to bring the one-electron energy level near $E_{F}$, so
that an electron can hop onto the dot. A short time after the pulse the electron will hop off the dot because it is above
$E_{F}$. We record the time at which this occurs, $\Delta t$, measured relative to the end of the gate pulse.  This process
is repeated and we make a histogram of the number of tunnel-off events vs. $\Delta t$; an exponential fit to this histogram
yields $\Gamma_{off}$. An analogous technique is used to measure $\Gamma_{on}$ when the one-electron state is well below
$E_{F}$.

To understand the $V_{g}$ dependence of $\Gamma$, we note that $\delta U_{2} = \alpha_{g U_{2}} \delta V_{g}$, and
$\alpha_{g U_{2}} < \alpha_{g E}$ because the three plunger gates are closer to the dot than they are to b2.  The
dependence of the tunnel rate through b2 on $V_{g}$ is thus very similar to its dependence on $V_{ds}$ (\totfig{fig3}).
Starting at the far left of \subfig{fig4}{a}, we see that $\Gamma_{off}$ decreases exponentially as $\Delta V_{g}$ is made
less negative: This is analogous to the $\Gamma_{off}$ data shown in the top panel of \subfig{fig2}{a} and \totfig{fig3}
(for $V_{ds} < 0$, where the electron tunnels off through b2). $\Gamma_{off}$ decreases rapidly at the 0 to 1 electron
transition as the ground state is brought below $E_{F}$.  Continuing into the 1 electron valley we see that $\Gamma_{on}$
decreases as $\Delta V_{g}$ is made more positive: This is analogous to the $\Gamma_{on}$ data shown in \totfig{fig3} (for
$V_{ds} > 0$ where the electron tunnels onto the dot through b2).  This decrease in $\Gamma_{on}$ is further evidence that
the tunneling is elastic because as the one-electron state is brought farther in energy below $E_{F}$ there are more
electrons that could tunnel onto the dot inelastically at energies closer to the top of the barrier. $\Gamma_{on}$, however,
decreases because elastic tunneling onto the dot happens at a lower energy relative to the barrier height.

We model the data shown in \subfig{fig4}{a} using the same equations as for the drain-source dependence (\eqn{eq:spinWKBoff}
and \eqn{eq:spinWKBon}) except with $\Gamma_{1,0}$ set to zero and $\delta V_{ds}$ replaced by $\Delta V_{g}$ in the
exponents.  The solid lines in \subfig{fig4}{a} are calculated from \eqn{eq:spinWKBoff} and \eqn{eq:spinWKBon} with these
adjustments and describe the data well.  Here again we take $\eta$ = 2, although the fit is improved with a smaller value
for $\eta$ \cite{Fujisawa:Bidirectional}. We note that the value for $\beta_{2}$ obtained from the $\Delta V_{g}$ dependence
is smaller than the value obtained from the $\delta V_{ds}$ dependence: This is expected because $\alpha_{g E}< \alpha_{ds
E}$.

Summarizing, we find that the dependence of $\Gamma_{on}$ and $\Gamma_{off}$ on $V_{ds}$ and $V_{g}$ is well described by
elastic tunneling at a rate set by the difference between the electron energy and the barrier height.

We are grateful to J. I. Climente, A. Bertoni, and G. Goldoni for calculations of acoustic phonon emission rates and discussions, to V. N. Golovach, L. Levitov, and D. Loss for discussions, and to I. J. Gelfand, T. Mentzel, V. Rosenhaus, and J. Schirmer for experimental help. This work was supported by the US Army Research Office under Contract W911NF-05-1-0062, by the National Science Foundation under Grant No. DMR-0353209, and in part by the NSEC Program of the National Science Foundation under Award Number PHY-0117795.


\end{document}